\documentclass[twocolumn,twoside]{IEEEtran}
\usepackage{amsmath,amssymb,amsthm,epsfig,dsfont,color,subfigure,empheq,graphicx}
\usepackage{enumerate,stfloats,url,algpseudocode,algorithm,wasysym,epstopdf,balance}
\usepackage[table]{xcolor}

\DeclareMathOperator{\diag}{diag}
\DeclareMathOperator{\bdiag}{bdiag}

\DeclareMathOperator{\real}{Re}
\DeclareMathOperator{\imag}{Im}

\DeclareMathOperator{\prox}{prox}

\newtheorem{proposition}{Proposition}

\newtheorem{corollary}{Corollary}

\theoremstyle{remark}\newtheorem{remark}{Remark}

\begin{document}
\title{Voltage Regulation Algorithms\\
for Multiphase Power Distribution Grids}

\author{
	Vassilis Kekatos,~\IEEEmembership{Member,~IEEE,}
	Liang Zhang,~\IEEEmembership{St. Member,~IEEE,}\\
	Georgios B. Giannakis,~\IEEEmembership{Fellow,~IEEE,} and
	Ross Baldick,~\IEEEmembership{Fellow,~IEEE}
\thanks{Work in this paper was supported by NSF grants 1423316, 1442686, 1508993, and 1509040. V. Kekatos is with the ECE Dept., Virginia Tech, Perry Str, Blacksburg, VA 24060, USA. L. Zhang and G. B. Giannakis are with the Digital Technology Center and the ECE Dept., University of Minnesota, Minneapolis, MN 55455, USA. R. Baldick is with the ECE Dept., University of Texas at Austin, TX 78712. Emails: kekatos@vt.edu, \{zhan3523,georgios\}@umn.edu, baldick@ece.utexas.edu}
}

\maketitle

\begin{abstract}
Time-varying renewable energy generation can result in serious under-/over-voltage conditions in future distribution grids. Augmenting conventional utility-owned voltage regulating equipment with the reactive power capabilities of distributed generation units is a viable solution. Local control options attaining global voltage regulation optimality at fast convergence rates is the goal here. In this context, novel reactive power control rules are analyzed under a unifying linearized grid model. For single-phase grids, our proximal gradient scheme has computational complexity comparable to that of the rule suggested by the IEEE 1547.8 standard, but it enjoys well-characterized convergence guarantees. Adding memory to the scheme results in accelerated convergence. For three-phase grids, it is shown that reactive injections have a counter-intuitive effect on bus voltage magnitudes across phases. Nevertheless, when our control scheme is applied to unbalanced conditions, it is shown to reach an equilibrium point. Yet this point may not correspond to the minimizer of a voltage regulation problem. Numerical tests using the IEEE 13-bus, the IEEE 123-bus, and a Southern California Edison 47-bus feeder with increased renewable penetration verify the convergence properties of the schemes and their resiliency to grid topology reconfigurations.
\end{abstract}


\section{Introduction}\label{sec:intro}
Voltage regulation, that is the task of maintaining bus voltage magnitudes within desirable levels, is critically challenged in modern distribution grids. The penetration of renewables, demand-response programs, and electric vehicles lead to time-varying active power injections and frequently reversing power flows. Utility-owned equipment conventionally employed for voltage regulation, such as tap-changing transformers and shunt capacitors, cannot react promptly and efficiently enough~\cite{BaldickWu90,Carvalho}. Hence, avoiding under-/over-voltage conditions requires advanced reactive power management solutions. To that end, the power electronics of PV inverters and storage devices offer a decentralized and fast-responding alternative~\cite{Rogers10,Turitsyn11}. 

A grid operator can engage the reactive power capabilities of distributed generation (DG) units to minimize power losses while satisfying voltage regulation constraints. Being an instance of the optimal power flow (OPF) problem, reactive power support can be solved using convex relaxation techniques~\cite{Jabr06,FCL}. Among other centralized approaches, inverter VAR control is solved using convex relaxation in \cite{FCL}, while a scheme relying on successive convex approximation is devised in~\cite{Deshmukh}. Distributed algorithms requiring communication across nodes have been proposed too. A distributed method based on convex relaxation has been developed in~\cite{Baosen14}. Upon modeling power losses as a quadratic function of reactive power injections, \cite{Saverio} pursues a consensus-type algorithm. Control rules based on approximate models are presented in~\cite{Turitsyn11}, and \cite{Markabi} developes a multi-agent scheme. Building on the radial structure of distribution grids, algorithms requiring communication only between adjacent nodes have been developed based on the alternating-direction method of multipliers (ADMM)~\cite{Chertkov}, \cite{QiuyuPeng}, \cite{GatsisN}.

To cater the scalability of DG units and the potential lack of communication infrastructure, local \emph{plug-and-play} schemes are highly desirable. Given that voltage magnitudes depend on grid-wide reactive injections, guaranteeing voltage regulation constraints may be hard to accomplish via purely localized algorithms. In such setups, reactive power management is usually relaxed to penalizing voltage magnitude deviations from a desired value and neglecting power losses. Reactive power injections are adjusted proportionally to the local voltage violations in~\cite{Robbins}; see also \cite{ZDGT13} for sufficient conditions guaranteeing its convergence. A similar local control strategy has been shown to minimize a modified voltage regulation cost in~\cite{FCL13}, while \cite{FZC15} proposes a subgradient-based algorithm. A control rule adjusting the inverter voltage output according to the reactive power flow is reported in~\cite{Dorfler}. The local control rules proposed in \cite{LQD14} maintain voltage magnitudes within the desired range under the presumption of unlimited reactive power support.

Most existing works build on a simplified single-phase grid model. Due to untransposed distribution lines and unbalanced loads though, the equivalent single-phase distribution network may not exist. Semidefinite programming and ADMM-based schemes have been applied in multiphase radial networks for power flow optimization \cite{HaozhuD}, \cite{PSCC14}. Nonetheless, no work discloses how inter-phase coupling affects bus voltage magnitudes across phases, or how local control schemes behave in unbalanced grids.

This work considers local reactive power control rules for voltage regulation in single- and multi-phase distribution grids. Our contribution is on four fronts. First, we provide a unified matrix-vector notation for approximate yet quite accurate multi-phase grid models (Section~\ref{sec:model}). Second, Section~\ref{sec:sp-alg} extends the work of~\cite{SGC15}. In \cite{SGC15}, we developed a reactive power control rule based on a proximal gradient scheme, and engineered modifications with superior convergence properties. Here, the options of unlimited reactive support and diagonal scaling are considered too. In particular, numerical results indicate that the convergence rates attained by using different step sizes across buses are still significantly lower than those achieved by our accelerated scheme. Third, using a linear approximation for unbalanced multi-phase grids, we reveal an interesting inter-phase coupling pattern across buses (Section~\ref{sec:mp-alg}). Recall that in single-phase grids, increasing the reactive power injection at any bus raises the nodal voltage magnitudes throughout the grid. In multi-phase grids on the contrary, injecting more reactive power into a bus of one phase could result in decreasing voltage magnitudes for the preceding in the positive-sequence ordering phase. It is finally shown that in unbalanced scenarios our reactive power control rule converges to an equilibrium point; yet this point does not necessarily correspond to the minimizer of a voltage regulation problem. Numerical tests on distribution feeders corroborate the convergence properties of the novel schemes, as well as their resiliency to topology reconfigurations (Section~\ref{sec:tests}).

Regarding \emph{notation}, lower- (upper-) case boldface letters denote column vectors (matrices), with the exception of line power flow vectors $(\mathbf{P},\mathbf{Q})$. Calligraphic symbols are reserved for sets. Symbol $^{\top}$ stands for transposition. Vectors $\mathbf{0}$, $\mathbf{1}$, and $\mathbf{e}_n$, are the all-zeros, all-ones, and the $n$-th canonical vectors, respectively. Symbol $\|\mathbf{x}\|_2$ denotes the $\ell_2$-norm of vector $\mathbf{x}$, while 
$\lambda_i(\mathbf{X})$ stands for the $i$-th largest eigenvalue of $\mathbf{X}$. Operator $\diag(\mathbf{x})$ defines a diagonal matrix having $\mathbf{x}$ on its diagonal, whereas $\diag(\mathbf{X})$ is the vector formed by the main diagonal of $\mathbf{X}$. Operator $\bdiag(\{\mathbf{X}_n\})$ defines a block diagonal matrix with $\mathbf{X}_n$'s as blocks. Operators $\real(z)$ and $\imag(z)$ return the real and imaginary part of complex number $z$. 
A matrix with non-negative entries is denoted by $\mathbf{X}\geq \mathbf{0}$, while $\mathbf{X}\succeq \mathbf{0}$ indicates a symmetric positive semidefinite matrix.
 

\section{Radial Distribution Grid Modeling}\label{sec:model}
Approximate models for single- and multi-phase grids are presented in this section.

\subsection{Modeling of Single-Phase Grids}\label{subsec:sp-model}
Distribution grids are typically operated as radial. A single-phase grid with $N+1$ buses can be modeled by a tree graph $\mathcal{T}=(\mathcal{N}_o,\mathcal{L})$ whose nodes $\mathcal{N}_o:=\{0,\ldots,N\}$ correspond to buses, and whose edges $\mathcal{L}$ correspond to distribution lines with cardinality $|\mathcal{L}|=N$. The feeder bus is indexed by $n=0$, whereas every non-feeder bus $n\in\mathcal{N}=\{1,\ldots,N\}$ has a unique parent bus denoted by $\pi_n$. Without loss of generality, nodes can be numbered such that $\pi_n < n$ for all $n\in\mathcal{N}$. For every bus $n\in\mathcal{N}_o$, let $v_n$ be its squared voltage magnitude, and $s_n=p_n+jq_n$ its complex power injection.

The distribution line connecting bus $n$ with its parent $\pi_n$ is indexed by $n$. For every line $n\in \mathcal{L}$, let $z_n=r_n+jx_n$, $\ell_n$, and $S_n=P_n+jQ_n$ be the line impedance, the squared current magnitude, and the complex power flow sent from the sending bus $\pi_n$, respectively. If $\mathcal{C}_n$ is the set of children buses for bus $n$, the grid can be modeled by the branch flow model~\cite{BW1}, \cite{BW2}
\begin{subequations}\label{eq:model}
\begin{align}
s_n&=\sum_{k\in\mathcal{C}_n}S_k  - S_n + \ell_n z_n \label{eq:ms}\\
v_n&=v_{\pi_n} - 2\real[z_n^{\star} S_n] +\ell_n |z_n|^2\label{eq:mv}\\
|S_n|^2&=v_{\pi_n} \ell_n\label{eq:ml}
\end{align}
\end{subequations}
for all $n\in \mathcal{N}$, and the initial condition $s_0=\sum_{k\in\mathcal{C}_0}S_k$. 

For notational brevity, collect all nodal quantities related to non-feeder buses in vectors $\mathbf{p}:=[p_1~\cdots~p_N]^{\top}$, $\mathbf{q}:=[q_1~\cdots~q_N]^{\top}$, and $\mathbf{v}:=[v_1~\cdots~v_N]^{\top}$. Similarly for lines, introduce vectors $\mathbf{r}:=[r_1~\cdots~r_N]^{\top}$, $\mathbf{x}:=[x_1~\cdots~x_N]^{\top}$, $\mathbf{P}:=[P_1~\cdots~P_N]^{\top}$, and $\mathbf{Q}:=[Q_1~\cdots~Q_N]^{\top}$. Define further the complex vectors $\mathbf{s}:=\mathbf{p}+j\mathbf{q}$, $\mathbf{z}:=\mathbf{r}+j\mathbf{x}$, and $\mathbf{S}:=\mathbf{P}+j\mathbf{Q}$. According to the approximate \emph{LinDistFlow} model, the grid is described by the linear equations~\cite{BW1}, \cite{BW2}
\begin{subequations}\label{eq:modelcompact}
\begin{align}
\mathbf{s}&=\mathbf{A}^{\top}\mathbf{S}\label{eq:mcs}\\
\mathbf{A}\mathbf{v}&=2\real[\mathbf{Z}^{\star}\mathbf{S}]  - \mathbf{a}_0 v_0\label{eq:mcv}
\end{align}
\end{subequations}
where $v_0$ is the squared voltage magnitude at the feeder; matrix $\mathbf{Z}$ is defined as $\mathbf{Z}:=\diag(\mathbf{z})$; and $\mathbf{A}$ is the \emph{reduced branch-bus incidence matrix} enjoying the following properties. 

\begin{proposition}[\cite{SGC15}]\label{pro:AF}
The negative of the reduced branch-bus incidence matrix $-\mathbf{A}$ and its inverse $\mathbf{F}:=-\mathbf{A}^{-1}$ satisfy:\\
(p1) they are both lower triangular with unit eigenvalues;\\
(p2) $\mathbf{F}\geq \mathbf{0}$; and $\mathbf{F}\mathbf{a}_0=\mathbf{1}_N$\\
where $\tilde{\mathbf{A}}=[\mathbf{a}_0~\mathbf{A}]$ is the full branch-bus incidence matrix.
\end{proposition}

Equation \eqref{eq:mcs} can be now expressed as $\mathbf{S}=-\mathbf{F}^{\top}\mathbf{s}$. Substituting the latter into \eqref{eq:mcv}, and premultiplying by $-\mathbf{F}$ yields $\mathbf{v}=2\real\left[\mathbf{F}\mathbf{Z}^*\mathbf{F}^{\top}\mathbf{s}\right] + v_0\mathbf{F}\mathbf{a}_0$. Proposition~\ref{pro:AF} and the properties of the real part operator provide~\cite{BW1}, \cite{FCL13}
\begin{equation}\label{eq:sp-model}
\mathbf{v}=\mathbf{R}\mathbf{p} + \mathbf{X}\mathbf{q} + v_0\mathbf{1}_N
\end{equation}
where $\mathbf{R}:=2\mathbf{F}\diag(\mathbf{r})\mathbf{F}^{\top}$ and $\mathbf{X}:=2\mathbf{F}\diag(\mathbf{x})\mathbf{F}^{\top}$; see also \cite{Saverio2} for a linear approximation model relating complex voltages to complex injections. It is well understood that in transmission grids and under regular load conditions and high reactance-to-resistance ratios, the nodal voltage magnitudes are approximately independent of active power injections. On the contrary, the approximate model of \eqref{eq:sp-model} confirms that voltage magnitudes in distribution grids depend significantly not only on reactive but active injections too. The dependence is roughly linear with the following properties.
\begin{remark}\label{re:triangular}
Although $\mathbf{F}$ is lower triangular, matrices $\mathbf{R}$ and $\mathbf{X}$ are generally full. Hence, local injection deviations affect voltage magnitudes globally.
\end{remark}

\begin{remark}\label{re:inverses}
Assuming $\mathbf{r}$ and $\mathbf{x}$ to be strictly positive, $\mathbf{R}$ and $\mathbf{X}$ are symmetric and strictly positive-definite by definition; cf.~\cite{FCL13} for a more elaborate proof. Further, their inverses are expressed as $\mathbf{R}^{-1} = \frac{1}{2}\mathbf{A}^{\top} \diag^{-1}(\mathbf{r})\mathbf{A}$ and $\mathbf{X}^{-1} = \frac{1}{2}\mathbf{A}^{\top} \diag^{-1}(\mathbf{x})\mathbf{A}$.
\end{remark}

\begin{remark}\label{re:M-mat}
Since $\mathbf{F}\geq \mathbf{0}$, it follows readily that $\mathbf{R}\geq \mathbf{0}$ and $\mathbf{X}\geq \mathbf{0}$. Hence, injecting more active or reactive power at any bus raises the voltage magnitudes in the entire grid.
\end{remark}

\subsection{Modeling of Multi-Phase Grids}\label{subsec:mp-model}
The focus shifts next to modeling multi-phase grids. For ease of exposition, it is first assumed that all buses are served by all three phases. For this reason, system variables are now represented by 3-dimensional vectors. Slightly abusing the notation used in Section~\ref{subsec:sp-model}, the complex voltages and the power injections at all phases of bus $n$ here are denoted by $\tilde{\mathbf{v}}_n$ and $\mathbf{s}_n$, respectively. Similarly, the complex currents and the complex power flows on all phases of line $n$ are represented by $\tilde{\mathbf{i}}_n$ and $\mathbf{S}_n$, respectively. The coupling across phases provides the multivariate version of Ohm's law:
\begin{equation}\label{eq:m-Ohm}
\tilde{\mathbf{v}}_n=\tilde{\mathbf{v}}_{\pi_n} - \mathbf{Z}_n \tilde{\mathbf{i}}_n
\end{equation}
where $\mathbf{Z}_n=\mathbf{Z}_n^{\top}$ is the phase impedance matrix for line $n$. If $\mathbf{S}_n$ is the flow on line $n$ seen from bus $\pi_n$, the flow received at bus $n$ is
\[\diag(\tilde{\mathbf{v}}_n)\tilde{\mathbf{i}}_n^{\star}=\mathbf{S}_n-\diag(\mathbf{Z}_n\tilde{\mathbf{i}}_n) \tilde{\mathbf{i}}_n^{\star}.\]
The multi-phase generalization of \eqref{eq:ms} reads
\begin{align}\label{eq:m-flow}
\mathbf{s}_n&=\sum_{k\in\mathcal{C}_n} \mathbf{S}_k  - \mathbf{S}_n + \diag(\mathbf{Z}_n\tilde{\mathbf{i}}_n) \tilde{\mathbf{i}}_n^{\star}
\end{align}
for all $n\in\mathcal{N}$. As advocated in \cite{PSCC14}, to obtain the multi-phase equivalent of the voltage drop equation \eqref{eq:mv}, multiply \eqref{eq:m-Ohm} by the complex Hermitian of each side, and maintain only the diagonal of the resultant matrix:
\begin{align}\label{eq:m-drop}
\diag\left(\tilde{\mathbf{v}}_n \tilde{\mathbf{v}}_n^H\right) =& \diag\left(\tilde{\mathbf{v}}_{\pi_n} \tilde{\mathbf{v}}_{\pi_n}^H\right) - 2\real\left[ \diag\left(\tilde{\mathbf{v}}_{\pi_n} \tilde{\mathbf{i}}_n^H \mathbf{Z}_n^H\right) \right]\nonumber\\
& + \diag\left(\mathbf{Z}_n \tilde{\mathbf{i}}_n \tilde{\mathbf{i}}_n^H \mathbf{Z}_n^H\right).
\end{align}
The full AC model for this multiphase grid is completed by replicating the definition for flows \eqref{eq:ml} on a per phase basis. Similar to single-phase grids, the model involves computationally inconvenient quadratic equations, but convex relaxations render the model tractable under appropriate conditions~\cite{PSCC14}.

Alternatively, one may resort to the simpler approximate model of \cite{PSCC14}. As for single-phase grids, because $\mathbf{Z}_n$'s have relatively small entries, the last summands in the right-hand sides (RHS) of \eqref{eq:m-flow} and \eqref{eq:m-drop} can be dropped. Regarding the second summand in the RHS of \eqref{eq:m-drop}, let us further assume that phase voltages are approximately balanced. By surrogating $\tilde{\mathbf{v}}_n$ by $\tilde{v}_n \boldsymbol{\alpha}$, where $\boldsymbol{\alpha}:=[1~\alpha~\alpha^2]^{\top}$ and $\alpha=e^{-j\frac{2\pi}{3}}$, the complex current vector $\tilde{\mathbf{i}}_n$ can be roughly expressed as
\begin{equation}\label{eq:iapprox}
\tilde{\mathbf{i}}_n^{\star} \approx \frac{1}{\tilde{v}_{\pi_n}} \diag(\mathbf{S}_n) \boldsymbol{\alpha}^{\star}
\end{equation}
and the outer product $\tilde{\mathbf{v}}_{\pi_n} \tilde{\mathbf{i}}_n^H$ can be thus replaced by $\boldsymbol{\alpha}\boldsymbol{\alpha}^H\diag(\mathbf{S}_n)$. Let $\mathbf{v}_n$ be the vector of per-phase squared voltage magnitudes for all three phases on bus $n$
\begin{equation*}
\mathbf{v}_n=\left[\begin{array}{c}v_n^a\\ v_n^b\\v_n^c\end{array}\right]=\diag(\tilde{\mathbf{v}}_n \tilde{\mathbf{v}}_n^H).
\end{equation*}
Then, equation \eqref{eq:m-drop} permits the approximation
\begin{align*}
\mathbf{v}_{\pi_n} - \mathbf{v}_n&= 2\real\left[\diag\left(\boldsymbol{\alpha} \boldsymbol{\alpha}^H\diag(\mathbf{S}_n)\mathbf{Z}_n^H\right)\right]
\end{align*}
where the argument inside the real operator simplifies as
\begin{align*}
\diag\left(\boldsymbol{\alpha} \left(\mathbf{Z}_n \diag(\mathbf{S}_n^{\star}) \boldsymbol{\alpha}\right)^H\right)
&=\diag(\boldsymbol{\alpha}) \mathbf{Z}_n^{\star} \diag(\mathbf{S}_n) \boldsymbol{\alpha}^{\star}\\
&=\diag(\boldsymbol{\alpha}) \mathbf{Z}_n^H \diag(\boldsymbol{\alpha}^{\star}) \mathbf{S}_n.
\end{align*}
The equalities follow from the properties of the $\diag$ operator: $\diag(\mathbf{x}\mathbf{y}^H)=\diag(\mathbf{x})\mathbf{y}^{\star}$ and $\diag(\mathbf{x})\mathbf{y}  = \diag(\mathbf{y})\mathbf{x}$. The approximate multi-phase model reads for all $n\in\mathcal{N}$
 \begin{subequations}\label{eq:m-amodel}
\begin{align}
\mathbf{s}_n&=\sum_{k\in\mathcal{C}_n} \mathbf{S}_k  - \mathbf{S}_n\label{eq:m-amodels}\\
\mathbf{v}_{\pi_n} - \mathbf{v}_n &=  2\real\left[\tilde{\mathbf{Z}}_n^* \mathbf{S}_n\right] \label{eq:m-amodelv}\\
\tilde{\mathbf{Z}}_n&:=\diag(\boldsymbol{\alpha}^*) \mathbf{Z}_n \diag(\boldsymbol{\alpha}).\label{eq:m-Zntilde}
\end{align}
\end{subequations}

Building on the grid model of \cite{PSCC14}, we express \eqref{eq:m-amodel} in a matrix-vector form and study the involved matrices to better understand voltage regulation schemes. Heed that system variables can be arranged either per bus or per phase. For example, the squared voltage magnitudes can be stacked as
\begin{equation}\label{eq:stacks}
\check{\mathbf{v}}:=\left[\begin{array}{c}
\check{\mathbf{v}}_{a}\\
\check{\mathbf{v}}_{b}\\
\check{\mathbf{v}}_{c}
\end{array}\right]\quad \textrm{or}\quad
\mathbf{v}:=\left[\begin{array}{c}
\mathbf{v}_1\\
\vdots\\
\mathbf{v}_{N}
\end{array}\right]
\end{equation}
where $\check{\mathbf{v}}_{\phi}:=[v_1^{\phi}~\ldots~v_N^{\phi}]^{\top}$ for $\phi\in\{a,b,c\}$. Likewise, injections can be arranged in $\check{\mathbf{s}}$ or $\mathbf{s}$, and flows in $\check{\mathbf{S}}$ or $\mathbf{S}$. It can be easily verified that the aforementioned vector pairs are related by
\begin{equation}\label{eq:permute}
\mathbf{v} = \mathbf{T}\check{\mathbf{v}},~\mathbf{s} = \mathbf{T}\check{\mathbf{s}}, ~\mathbf{S} = \mathbf{T}\check{\mathbf{S}}
\end{equation}
for a common permutation matrix $\mathbf{T}$ compactly expressed as
\begin{equation}\label{eq:T}
\mathbf{T}:=\left[\begin{array}{c}
\mathbf{I}_3 \otimes \mathbf{e}_1^{\top}\\
\vdots\\
\mathbf{I}_3 \otimes \mathbf{e}_N^{\top}
\end{array}\right]
\end{equation}
where $\mathbf{e}_n$ is the $n$-th column of $\mathbf{I}_N$. Being a permutation matrix, $\mathbf{T}$ satisfies $\mathbf{T}^{-1}=\mathbf{T}^{\top}$. Algebraic manipulations postponed for the Appendix show that voltage magnitudes in multi-phase grids are related to nodal injections as follows:

\begin{proposition}\label{pro:B}
Based on \eqref{eq:m-amodel}, it holds that
\begin{equation}\label{eq:mp-model}
\mathbf{v} = \mathbf{R}\mathbf{p} + \mathbf{X}\mathbf{q} + v_0\mathbf{1}_{3N}
\end{equation}
where the involved matrices are defined as 
\begin{subequations}\label{eq:mp-all}
\begin{align}
\mathbf{R}&:=2\mathbf{M} \bdiag (\{\real [ \tilde{\mathbf{Z}}_n] \})  \mathbf{M}^{\top}\label{eq:mp-R}\\
\mathbf{X}&:=2\mathbf{M} \bdiag (\{\imag [ \tilde{\mathbf{Z}}_n] \})  \mathbf{M}^{\top},~\textrm{and}\label{eq:mp-X}\\
\mathbf{M}&:=\mathbf{T}(\mathbf{I}_3\otimes \mathbf{F}) \mathbf{T}^{\top}\label{eq:mp-M}
\end{align}
\end{subequations}
and matrices $\tilde{\mathbf{Z}}_n$ have been defined in \eqref{eq:m-Zntilde}.
\end{proposition}
According to \eqref{eq:m-amodels}, power injections and flows are approximately decoupled across phases. Nonetheless, Proposition~\ref{pro:B} asserts that squared voltage magnitudes depend on power injections from all phases. Building on the approximate models of this section, voltage regulation schemes are developed next.

\section{Schemes for Single-Phase Grids}\label{sec:sp-alg}
Posing voltage regulation as an optimal power flow instance leads to a constrained optimization problem. Given that voltage regulation constraints couple reactive injections across the grid, developing localized solutions becomes challenging. To derive such solutions, the voltage regulation goal is relaxed here and posed as the generic minimization problem
\begin{equation}\label{eq:gvr}
\min_{\mathbf{q}\in \mathcal{Q}} f(\mathbf{q}) + c(\mathbf{q})
\end{equation}
where $f(\mathbf{q})$ is the cost of squared voltage magnitudes $\mathbf{v}$ deviating from their nominal value $v_0\mathbf{1}$; $c(\mathbf{q})$ models the potential cost for reactive power compensation; and $\mathcal{Q}$ is the feasible set of reactive injections. Since reactive power injections by DG inverters can be adjusted in real-time, whereas utility-owned voltage regulating equipment responds typically at a slower pace (e.g., every few minutes or hourly), the latter choice is assumed fixed to a value and it will not be considered here. Particular instances of the generic setup in \eqref{eq:gvr} are instantiated next for single-phase grids.

\subsection{Unconstrained Reactive Support}\label{subsec:sp-uncon}
A viable voltage deviation cost is $f_1(\mathbf{q}):=\frac{1}{2}\|\mathbf{v}-v_0\mathbf{1}\|_2^2$. This cost function tries to keep squared voltage magnitudes close to the nominal value $v_0$. Assuming $c(\mathbf{q})=0$ and $\mathcal{Q}=\mathbb{R}^N$, problem \eqref{eq:gvr} boils down to the unconstrained quadratic program
\begin{equation}\label{eq:sp-gvr-noc}
\min_{\mathbf{q}}\tfrac{1}{2}\|\mathbf{R}\mathbf{p} + \mathbf{X}\mathbf{q}\|_2^2.
\end{equation}
Localized voltage regulation schemes assuming unlimited reactive power support have also been considered in \cite{ZDGT13} and \cite{LQD14}. Obviously, since $\mathbf{X}$ is invertible, problem \eqref{eq:sp-gvr-noc} has the unique minimizer $\mathbf{q}^{\star}=-\mathbf{X}^{-1}\mathbf{R}\mathbf{p}$ that achieves perfect voltage regulation $\mathbf{v}(\mathbf{q}^{\star})=v_0\mathbf{1}$. Finding $\mathbf{q}^{\star}$ requires knowing the active injections $\mathbf{p}$ over all buses. Using the structure of $\mathbf{R}$ and $\mathbf{X}$, vector $\mathbf{q}^{\star}$ can be alternatively expressed as
\begin{equation}\label{eq:closed-form}
\mathbf{q}^{\star} = \mathbf{A}^{\top} \diag\left(\left\{\frac{r_n}{x_n}\right\}\right) \mathbf{P}.
\end{equation}
The entry $q_n^{\star}$ is a linear combination of the active powers flowing in and out of bus $n$ with the related $r_n/x_n$ ratios as coefficients. Although $P_n$ denotes the power flow seen from the sending end of line $n$, the receiving end will measure approximately $-P_n$ due to the small loss assumption. The minimizer \eqref{eq:closed-form} can be found in a localized way only if bus $n$ measures power flows on incident lines. 

Alternatively, a gradient descent scheme would iteratively update reactive injections over time $t$ as
\begin{equation}\label{eq:sp-gd}
\mathbf{q}^{t+1}=\mathbf{q}^t-\mu \mathbf{g}^t
\end{equation}
where $\mu>0$ is a step size, and $\mathbf{g}^t=\mathbf{X}^{\top}(\mathbf{R}\mathbf{p} + \mathbf{X}\mathbf{q}^t)=\mathbf{X}(\mathbf{v}^t-v_0\mathbf{1})$ is the gradient of $f_1(\mathbf{q})$ at $\mathbf{q}^t$. Unfortunately, such a scheme cannot be implemented in a localized fashion. However, the next proposition proved in the Appendix asserts that the rule
\begin{equation}\label{eq:sp-agd}
\mathbf{q}^{t+1}=\mathbf{q}^t-\mu (\mathbf{v}^t-v_0\mathbf{1})
\end{equation}
converges to $\mathbf{q}^{\star}$ for an appropriately small $\mu$.
\begin{proposition}\label{pro:sp-descent}
If $\mu\in \left(0,2\lambda_{\min}(\mathbf{X})/\lambda_{\max}^2(\mathbf{X})\right)$, the rule of \eqref{eq:sp-agd} converges to the minimizer of \eqref{eq:sp-gvr-noc}.
\end{proposition}

The descent rule in \eqref{eq:sp-agd} scaled by a diagonal matrix has been shown to converge for a more detailed model in~\cite{ZDGT13}.

\subsection{Constrained Reactive Support}\label{subsec:sp-con}

Solving \eqref{eq:sp-gvr-noc} may be of little practical use: Distributed generation units may not be installed on every bus and their reactive power resources are limited. Moreover, the power electronics found on a PV at bus $n$ have finite apparent power capability $s_n$, which limits $q_n$ depending on the current active injection (solar generation) as $p_n^2+q_n^2\leq s_n^2$. In reality, $\mathbf{q}$ is constrained to lie within the time-varying but known box
\[\mathcal{Q}:=\{\mathbf{q}:|q_n|\leq \overline{q}_n:=\sqrt{s_n^2-p_n^2}\geq 0~\forall n\}.\]
Buses with reactive power support can be obviously modeled by selecting their associated limits as $\overline{q}_n=0$.

In this practically pertinent setup where the voltage regulation problem in \eqref{eq:gvr} is constrained as $\mathbf{q}\in\mathcal{Q}$, one could try implementing the projected version of \eqref{eq:sp-agd}, that is
\begin{equation}\label{eq:sp-agd-c}
\mathbf{q}^{t+1}=\left[\mathbf{q}^t-\mu (\mathbf{v}^t-v_0\mathbf{1})\right]_{\mathcal{Q}}
\end{equation} 
where $[\mathbf{x}]_{\mathcal{Q}}:=\arg\min_{\mathbf{z}\in \mathcal{Q}} \|\mathbf{x}-\mathbf{z}\|_2$ denotes the projection operator on $\mathcal{Q}$. Unfortunately, this seemingly meaningful control rule is not guaranteed to converge~\cite{BeTs97}.

A localized voltage regulation scheme can be obtained via a different voltage deviation cost $f(\mathbf{q})$. As advocated in \cite{FCL13}, a meaningful choice is the cost
\begin{align}\label{eq:sp-f2}
f_2(\mathbf{q})=\tfrac{1}{2}\|\mathbf{R}\mathbf{p}+\mathbf{X}\mathbf{q}\|_{\mathbf{X}^{-1}}^2
\end{align}
with the rotated norm defined as $\|\mathbf{x}\|_{\mathbf{B}}^2:=\mathbf{x}^{\top}\mathbf{B}\mathbf{x}$ for $\mathbf{B}\succ \mathbf{0}$.  As proved in \cite{SGC15}, the cost in \eqref{eq:sp-f2} is equivalent to $f_2(\mathbf{q})=\frac{1}{4}\sum_{n=1}^N \frac{(v_{\pi_n}-v_n)^2}{x_n}$. Although minimizing $f_2(\mathbf{q})$ over $\mathbf{q}\in \mathcal{Q}$ penalizes scaled voltage magnitude deviations between adjacent buses, obviously, it does not guarantee that voltages will lie within any desired range. Nevertheless, $f_2(\mathbf{q})$ has the important feature that its gradient
\begin{equation}\label{eq:sp-grad}
\nabla f_2(\mathbf{q})=\mathbf{R}\mathbf{p}+\mathbf{X}\mathbf{q}=\mathbf{v}(\mathbf{q})-v_0\mathbf{1}
\end{equation}
equals the deviation of squared voltage magnitudes from the nominal, and its $n$-th entry can be measured locally at bus $n$.

\begin{figure}[t]
\centering
\includegraphics[scale=0.35]{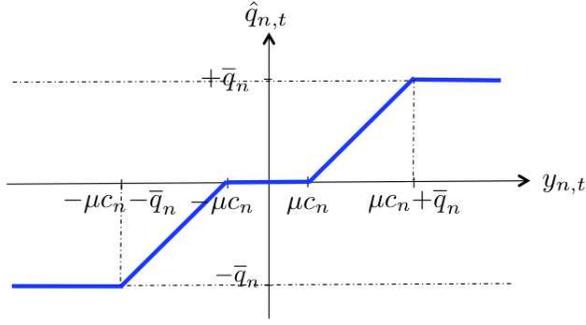}
\caption{The reactive control rule of \eqref{eq:closed} for $y_n^t=q_n^t -\mu(v_n^t-v_0)$.}
\label{fig:rule}
\end{figure}

To deter engaging PV power inverters for negligible voltage deviations, a reactive power compensation cost $c(\mathbf{q})$ should be also considered. Given that negative and positive reactive power injections are equally important, a reasonable option for voltage regulation would be solving the problem 
\begin{equation}\label{eq:gvr-f2}
\min_{\mathbf{q}\in \mathcal{Q}}~h_2(\mathbf{q})=f_2(\mathbf{q}) + c_2(\mathbf{q})
\end{equation}
where $c_2(\mathbf{q}):=\sum_{n=1}^N c_n |q_n|$. Again, due to the strong convexity of $f_2(\mathbf{q})$, problem \eqref{eq:gvr-f2} has a unique minimizer in $\mathcal{Q}$. As shown in \cite{SGC15}, the minimizer of \eqref{eq:gvr-f2} can be found via simple proximal gradient iterations: At iterate $t$, each bus $n$ measures the quantity
\begin{equation}\label{eq:sp-yn}
y_n^t:=q_n^t -\mu(v_n^t-v_0)
\end{equation}
for a step size $\mu>0$. Voltage magnitude deviations $v_n^t-v_0$ are assumed to be acquired without noise. It then updates its reactive power injection according to the control rule
\begin{equation}\label{eq:closed}
q_{n}^{t+1} :=\mathcal{S}_{\mu}^{\overline{q}_n}(y_n^t,c_n).
\end{equation}
The operator $\mathcal{S}_{\mu}^{\overline{q}}(y,c)$ is defined as
\begin{equation}\label{eq:rule}
\mathcal{S}_{\mu}^{\overline{q}}(y,c) :=\left\{ \begin{array}{ll}
+\overline{q}		&,~	y> \overline{q} + \mu c\\
 y - \mu c		&,~	\mu c < y \leq \overline{q} + \mu c\\
0						&,~	-\mu c \leq y \leq \mu c\\
 y + \mu c		&,~	 -\overline{q}- \mu c \leq y < -\mu c\\
-\overline{q}		&,~	y< -\overline{q} - \mu c
\end{array}\right.
\end{equation}
and is shown in Figure~\ref{fig:rule}. Apparently, in the absence of reactive power compensation cost, that is when $c=0$, the operator $\mathcal{S}_{\mu}^{\overline{q}}(y,0)$ simply projects $y$ onto $[-\overline{q},\overline{q}]$.

If the step size is selected as $\mu\in\left(0,2\lambda_{\max}^{-1}(\mathbf{X})\right)$, the control rule of \eqref{eq:sp-yn}--\eqref{eq:closed} converges to the minimizer of \eqref{eq:gvr-f2} \cite{proximal}. Actually, with the optimal step size $\mu=\lambda_{\max}^{-1}(\mathbf{X})$, the convergence rate is linear, but proportional to the condition number $\kappa(\mathbf{X}):=\lambda_{\max}(\mathbf{X})/\lambda_{\min}(\mathbf{X})$ of matrix $\mathbf{X}$. It is worth stressing that $\kappa(\mathbf{X})$ can be relatively large: the Southern California Edison 47-bus grid and the IEEE 34-bus benchmark exhibit $\kappa(\mathbf{X})=1.6\times  10^4$ and $5.5\times 10^4$, respectively~\cite{FCL},~\cite{PSTCA}. Ways to improve the convergence speed are reviewed next.

\subsection{Accelerating Voltage Regulation Schemes}\label{subsec:sp-acc}
A way to speed up proximal gradient schemes is via diagonal scaling. In detail, the reactive injection variables $\mathbf{q}$ can be transformed to $\breve{\mathbf{q}}$ as $\mathbf{q}=\mathbf{D}^{1/2}\breve{\mathbf{q}}$ for a diagonal matrix $\mathbf{D}\succ \mathbf{0}$. Instead of \eqref{eq:gvr-f2}, we can solve the equivalent problem
\begin{equation*}
\min_{\mathbf{D}^{1/2}\breve{\mathbf{q}} \in \mathcal{Q}} h_2(\mathbf{D}^{1/2}\breve{\mathbf{q}})
\end{equation*}
through the proximal gradient method. The update for the transformed variable $\breve{\mathbf{q}}^{t+1}$ is found as the solution to
\begin{equation*}
\min_{\mathbf{D}^{1/2}\breve{\mathbf{q}}\in \mathcal{Q}}  c_2(\mathbf{D}^{1/2}\breve{\mathbf{q}})+\tfrac{1}{2\mu}\|\breve{\mathbf{q}}- (\breve{\mathbf{q}}^t-\mu\mathbf{D}^{1/2}\nabla f_2(\mathbf{D}^{1/2}\breve{\mathbf{q}}^t))\|_2^2
\end{equation*}
and the convergence rate now depends on  $\kappa \left(\mathbf{D}^{1/2}\mathbf{X}\mathbf{D}^{1/2}\right)$. Translating the iterations back to the original variables $\mathbf{q}$ yields
\begin{equation}\label{eq:sp-pg-scaled}
\mathbf{q}^{t+1}{:=}\arg\min_{\mathbf{q}\in \mathcal{Q}} c_2(\mathbf{q}) +\tfrac{1}{2\mu}\|\mathbf{q}- (\mathbf{q}^t-\mu\mathbf{D}\nabla f_2(\mathbf{q}^t))\|_{\mathbf{D}^{-1}}^2
\end{equation}
which is separable across buses as
\begin{equation}\label{eq:sp-prox-local}
q_n^{t+1}=\arg\min_{\underline{q}_n\leq q_n\leq \bar{q}_n} ~ d_n c_n|q_n| +\frac{1}{2\mu} \left( q_n -u_n^t \right)^2
\end{equation}
where $u_n^t:=q_n^t-\mu d_n (v_n^t-v_0)$. It can be easily verified that the minimizer of \eqref{eq:sp-pg-scaled} is provided by the control rule 
\begin{equation}\label{eq:closed2}
q_{n}^{t+1} :=\mathcal{S}_{\mu d_n}^{\overline{q}_n}(u_n^t,d_nc_n)
\end{equation}
where $d_n$ is the $n$-th diagonal entry of the scaling matrix $\mathbf{D}$. The control rule in \eqref{eq:closed2} corresponds to the rule of \eqref{eq:sp-yn}--\eqref{eq:closed} with the step size $\mu$ adjusted to $\mu d_n$ per bus $n$. The scaling matrix $\mathbf{D}$ could be selected to minimize $\kappa \left(\mathbf{D}^{1/2}\mathbf{X}\mathbf{D}^{1/2}\right)$. A relatively simple choice for $\mathbf{D}$ is to assign vector $\diag(\mathbf{X})$ on the main diagonal of $\mathbf{D}^{-1}$. This option sets the diagonal entries of $\mathbf{D}^{1/2}\mathbf{X}\mathbf{D}^{1/2}$ to unity. To meet faster voltage regulation rates, the accelerated voltage regulation scheme of \cite{SGC15} is reviewed and simplified next.

In \cite{SGC15}, we derived an accelerated proximal gradient scheme by adapting Nesterov's method~\cite{Nesterov}. The corresponding control rule was shown to be a simple modification of \eqref{eq:closed} (see \cite{SGC15} for details):
\begin{equation}\label{eq:closed3}
q_{n}^{t+1} :=\mathcal{S}_{\mu}^{\overline{q}_n}(\tilde{y}_n^t,c_n)
\end{equation} 
where variable $\tilde{y}_n^t$ is updated using the $y_n^t$ from \eqref{eq:sp-yn} as
\begin{equation}\label{eq:sp-yns}
\tilde{y}_n^t:=(1+\beta_t)y_n^t-\beta_t y_n^{t-1}
\end{equation}
with $\beta_t=\frac{t-1}{t+2}$ for all $t\geq 1$. Compared to \eqref{eq:closed}, the control rule in \eqref{eq:closed3} introduces memory in calculating $\tilde{y}_n^t$ as a linear combination of $y_n^t$ and $y_n^{t-1}$. The linear combination coefficients depend on the time-varying $\beta_t$ that converges to 1. Note that the formula for $\beta_t$ has been simplified from the one used in~\cite[Eqs.~(20)-(21)]{SGC15}.

If the step size is selected as $\mu=\lambda_{\max}^{-1}(\mathbf{X})$ and the sequence $\beta_t$ is reset to zero every $2\sqrt{\kappa(\mathbf{X})}$ iterations, the reactive control rule of \eqref{eq:closed3} converges linearly to an $\epsilon$-optimal cost value within $-\frac{2\log \epsilon}{\log 2} \sqrt{\kappa(\mathbf{X})}$ iterations. For grids with high $\kappa(\mathbf{X})$, this modified scheme offers accelerated convergence. Numerical tests show that even without resetting the sequences and without knowing precisely $\lambda_{\max}(\mathbf{X})$, the rule in \eqref{eq:closed3} offers superior performance over both rules \eqref{eq:closed} and \eqref{eq:closed2}.

\begin{remark}
The IEEE 1547.8 standard suggests the following reactive power injection rule for the DG at bus $n$~\cite{IEEE1547}
\begin{equation}\label{eq:1547}
q_n^{t+1}=S_n(v_0-v_n^t)
\end{equation}
where the function $S_n$ is an increasing piecewise linear function similar to the one shown in Figure~\ref{fig:rule}. Comparing \eqref{eq:1547} with the control rules of \eqref{eq:closed} or \eqref{eq:closed2}, suggests that all control rules have similar computational requirements. Nevertheless, the rule in \eqref{eq:1547} can exhibit oscillations as reported in \cite{FZC15}, while the schemes presented here exhibit well-understood convergence properties.
\end{remark}

\section{Schemes for Multi-Phase Grids}\label{sec:mp-alg}
Distributed generation and demand-response programs can lead to unbalanced conditions. This section aims at generalizing the schemes of Section~\ref{sec:sp-alg} for multi-phase grids. The problem of voltage regulation in multi-phase grids can be posed as in \eqref{eq:gvr}; yet now the dependence of squared voltage magnitudes on nodal injections is governed by the model in \eqref{eq:mp-model}. Before devising voltage regulation schemes, critical properties of the involved matrices are studied first.

Let us focus on the $3\times 3$ complex matrices $\mathbf{\tilde{Z}}_n$ defined in \eqref{eq:m-amodel}. Let $z_n^{ij}=r_n^{ij}+jx_n^{ij}$ be the $(i,j)$-th entry of $\mathbf{Z}_n$. From the symmetry of $\mathbf{Z}_n$ and the identity $\alpha^2=\alpha^*=-\frac{1}{2}+j\frac{\sqrt{3}}{2}$, matrix $\mathbf{\tilde{Z}}_n$ becomes
\begin{equation*}
\mathbf{\tilde{Z}}_n=\diag(\boldsymbol{\alpha}^*) \mathbf{Z}_n \diag(\boldsymbol{\alpha})=\left[\begin{array}{ccc}
z_n^{11} & \alpha^* z_n^{12} & \alpha z_n^{13}\\
\alpha z_n^{21} & z_n^{22} & \alpha^* z_n^{23}\\
\alpha^* z_n^{31} & \alpha z_n^{32} & z_n^{33}
\end{array}\right].
\end{equation*}
Therefore, matrix $\imag[\mathbf{\tilde{Z}}_n]$ can be decomposed as
\begin{subequations}\label{eq:ImZntilde}
\begin{align}
\imag[\mathbf{\tilde{Z}}_n] &= \tilde{\mathbf{X}}_n + \tilde{\mathbf{R}}_n \label{eq:ImZntilde-d}\\
\tilde{\mathbf{X}}_n&=\frac{1}{2}\left[\begin{array}{ccc}
2x_n^{11} & - x_n^{12} & -x_n^{13}\\
-x_n^{12} & 2x_n^{22} & -x_n^{23}\\
-x_n^{13} &  -x_n^{23}& 2x_n^{33}
\end{array}\right] \label{eq:ImZntilde-X}\\
\tilde{\mathbf{R}}_n&= \frac{\sqrt{3}}{2}
\left[\begin{array}{ccc}
0 & - r_n^{12} &  r_n^{13}\\
r_n^{12} & 0 & -r_n^{23}\\
-r_n^{13} & r_n^{23}& 0
\end{array}\right] \label{eq:ImZntilde-R}
\end{align}
\end{subequations}
where $\tilde{\mathbf{X}}_n$ is a symmetric matrix $(\tilde{\mathbf{X}}_n=\tilde{\mathbf{X}}_n^{\top})$ associated to the reactive part of $\mathbf{Z}_n$, and $\tilde{\mathbf{R}}_n$ is an anti-symmetric matrix $(\tilde{\mathbf{R}}_n=-\tilde{\mathbf{R}}_n^{\top})$ depending on the resistive part of $\mathbf{Z}_n$. Using \eqref{eq:ImZntilde}, the next fact holds:
\begin{proposition}\label{pro:Zn}
If $\tilde{\mathbf{X}}_n$ is strictly diagonally dominant with positive diagonal entries, then $\imag[\mathbf{\tilde{Z}}_n]\succ \mathbf{0}$.
\end{proposition}

To prove Proposition~\ref{pro:Zn}, it suffices to show that the symmetric component of $\imag[\mathbf{\tilde{Z}}_n]$ is strictly positive-definite. If mutual and self-reactances satisfy $2x_n^{ii}>\sum_{j\neq i} |x_n^{ij}|$ for all $i$, then $\tilde{\mathbf{X}}_n$ is diagonally dominant with positive diagonal entries, and thus, $\tilde{\mathbf{X}}_n\succ \mathbf{0}$. Due to the structure of distribution lines, the aforementioned inequalities are satisfied in general.

The decomposition in \eqref{eq:ImZntilde} carries over to $\mathbf{X}$ in \eqref{eq:mp-X} as:
\begin{equation}\label{eq:Xdec}
\mathbf{X}=\mathbf{X}_x + \mathbf{X}_r
\end{equation}
where $\mathbf{X}_x:=2\mathbf{M} \bdiag (\{ \tilde{\mathbf{X}}_n\})\mathbf{M}^{\top}$ is symmetric, and $\mathbf{X}_r := 2\mathbf{M} \bdiag (\{ \tilde{\mathbf{R}}_n\})\mathbf{M}^{\top}$ is anti-symmetric. Because $\mathbf{M}$ is invertible, matrix $\mathbf{X}$ is positive-definite if and only if matrix $\bdiag(\{\tilde{\mathbf{X}}_n\})$ is, hence leading to the corollary:
\begin{corollary}\label{co:mp-X}
If $\imag [\mathbf{Z}_n]$ is strictly diagonally dominant with positive diagonal entries for all $n$, then $\mathbf{X}\succeq \mathbf{0}$.
\end{corollary}

\subsection{Inter-Phase Coupling}\label{subsec:coupling}
We next elaborate on how bus voltage magnitudes are affected by reactive power injections. According to \eqref{eq:mp-model}, vector $\check{\mathbf{v}}$ is linearly related to reactive injections $\check{\mathbf{q}}$ via the matrix 
\begin{align}\label{eq:Xcheck}
\check{\mathbf{X}}&:=\mathbf{T}^{\top}\mathbf{X}\mathbf{T}\\
&=2(\mathbf{I}_3 \otimes \mathbf{F})\mathbf{T}^{\top} \bdiag (\{\imag [ \tilde{\mathbf{Z}}_n] \}) \mathbf{T} (\mathbf{I}_3 \otimes \mathbf{F}^{\top}).\nonumber
\end{align}
The effect of reactive power injection $\check{\mathbf{q}}_j$ to the squared voltage magnitude $\check{\mathbf{v}}_i$ is described by the $(i,j)$ entry of $\check{\mathbf{X}}$. Let entry $i$ correspond to phase $\phi_i$ of bus $n_i$, and entry $j$ to phase $\phi_j$ of bus $n_j$. It can be verified that
\begin{align*}
\check{\mathbf{X}}_{i,j}&=(\mathbf{e}_{\phi_i}\otimes \mathbf{e}_{n_i})^{\top}\check{\mathbf{X}}(\mathbf{e}_{\phi_j}\otimes \mathbf{e}_{n_j})\\
&=(\mathbf{e}_{\phi_i}\otimes \mathbf{f}_{n_i})^{\top}\mathbf{T}^{\top} \bdiag (\{\imag [ \tilde{\mathbf{Z}}_n] \}) \mathbf{T}(\mathbf{e}_{\phi_j}\otimes \mathbf{f}_{n_j})\nonumber
\end{align*}
where $\mathbf{f}_k^{\top}$ is the $k$-th row of matrix $\mathbf{F}$. By the definition of $\mathbf{T}$ in \eqref{eq:T}, the products $\mathbf{T}(\mathbf{e}_{\phi_i}\otimes \mathbf{f}_{n_i})$ and $\mathbf{T}(\mathbf{e}_{\phi_j}\otimes \mathbf{f}_{n_j})$ can be expressed as $\mathbf{f}_{n_i}\otimes \mathbf{e}_{\phi_i}$ and $\mathbf{f}_{n_j}\otimes \mathbf{e}_{\phi_j}$, respectively. Exploiting the structure of $\bdiag (\{\imag [ \tilde{\mathbf{Z}}_n] \})$ and since $\mathbf{F}$ is lower triangular, the entry $\check{\mathbf{X}}_{i,j}$ can be expressed as
\begin{equation}\label{eq:Xcheckij2}
\check{\mathbf{X}}_{i,j}=\sum_{k=1}^{\min\{n_i,n_j\}} \imag [ \tilde{\mathbf{Z}}_k] _{\phi_i,\phi_j} \mathbf{F}_{n_i,k} \mathbf{F}_{n_j,k}.
\end{equation}
Recall that $\mathbf{F}$ has non-negative entries, while for overhead distribution lines the parameters $x_n^{\phi_i\phi_j}$ and $r_n^{\phi_i\phi_j}$ appearing in \eqref{eq:ImZntilde-X}--\eqref{eq:ImZntilde-R} are typically non-negative. According to \eqref{eq:Xcheckij2}, three cases can be distinguished:\\
\hspace*{1em}\textbf{(c1)} If $\phi_i=\phi_j$, $\check{\mathbf{X}}_{i,j}=\sum_{k=1}^{\min\{n_i,n_j\}} x_k^{\phi_i\phi_i} \mathbf{F}_{n_i,k} \mathbf{F}_{n_j,k}>0$. Thus, as in single-phase grids, injecting more reactive power into a bus raises voltages at all buses in the \emph{same phase}.\\
\hspace*{1em}\textbf{(c2)} When $(\phi_i,\phi_j)\in \{(a,b),(b,c),(c,a)\}$, it follows that $\check{\mathbf{X}}_{i,j}=-\tfrac{1}{2}\sum_{k=1}^{\min\{n_i,n_j\}} \left(x_k^{\phi_i\phi_i}+\sqrt{3}r_k^{\phi_i\phi_j}\right) \mathbf{F}_{n_i,k} \mathbf{F}_{n_j,k}<0$. Thus, injecting more reactive power into a bus decreases the voltage magnitudes at all buses of the \emph{preceding phase} in the positive sequence ordering.\\
\hspace*{1em}\textbf{(c3)} If $(\phi_i,\phi_j)\in \{(a,c),(b,a),(c,b)\}$, then $\check{\mathbf{X}}_{i,j}=-\tfrac{1}{2}\sum_{k=1}^{\min\{n_i,n_j\}} \left(x_k^{\phi_i\phi_i}-\sqrt{3}r_k^{\phi_i\phi_j}\right) \mathbf{F}_{n_i,k} \mathbf{F}_{n_j,k}$. Thus, the effect of reactive power injections into one phase to the voltage magnitudes of the \emph{following phase} depends on the differences $x_k^{\phi_i\phi_i}-\sqrt{3}r_k^{\phi_i\phi_j}$ appearing in the last sum. Actually, if every bus serves all phases and since $|x_k^{\phi_i\phi_i}-\sqrt{3}r_k^{\phi_i\phi_j}|\leq x_k^{\phi_i\phi_i}+\sqrt{3}r_k^{\phi_i\phi_j}$, the effect of one phase to the following phase is less significant than its effect on the previous phase.

\begin{figure}[t]
\centering
\includegraphics[scale=0.6]{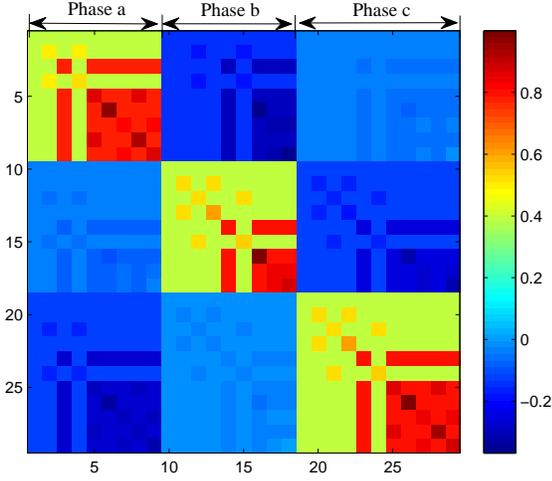}
\vspace*{-2.5em}
\caption{Matrix $\check{\mathbf{X}}$ (normalized to unity maximum entry) relating voltage magnitudes $\check{\mathbf{v}}$ to reactive power injections $\check{\mathbf{q}}$ 
for the IEEE 13-bus grid benchmark depicts the coupling across phases.}
\label{fig:pattern}
\end{figure}

An illustration of the influence patterns across phases is shown in Figure~\ref{fig:pattern}. Evidenced by the previous analysis and Fig.~\ref{fig:pattern}, decreasing reactive injections in phase $b$ to cater over-voltage conditions on phase $b$ would aggravate possible over-voltage problems on phase $a$. In this context, voltage regulation becomes even more challenging in multi-phase grids.

\subsection{Unconstrained Reactive Support}\label{subsec:mp-uncon}
Similar to single-phase grids, let us first consider the simplest voltage regulation scenario: The voltage deviation cost is $f_1(\mathbf{q})=\tfrac{1}{2}\|\mathbf{v}-v_0\mathbf{1}_{3N}\|_2^2$, there is no reactive power compensation cost $c(\mathbf{q})$, and reactive power is unconstrained:
\begin{equation}\label{eq:mp-gvr-noc}
\min_{\mathbf{q}}\tfrac{1}{2}\|\mathbf{R}\mathbf{p} + \mathbf{X}\mathbf{q}\|_2^2.
\end{equation}
Again, the positive-definiteness of $\mathbf{X}$ [cf.~Corollary.~\ref{co:mp-X}] guarantees the uniqueness of the minimizer in \eqref{eq:mp-gvr-noc}. Contrary to the single-phase case though, the minimizer of \eqref{eq:mp-gvr-noc} cannot be found locally even if bus $n$ measures all phase flows on its incident lines [cf.~\eqref{eq:closed-form}] because of the structure of $\mathbf{R}$ and $\mathbf{X}$.

Similar to single-phase grids, a gradient descent scheme cannot be implemented locally. However, the simple update
\begin{equation}\label{eq:mp-rule}
\mathbf{q}^{t+1}=\mathbf{q}^t -\mu (\mathbf{v}^t-v_0\mathbf{1})
\end{equation}
still constitutes a convergent rule:
\begin{proposition}\label{pro:mp-descent}
If $\mu\in \left(0,\frac{2\lambda_{\min}(\mathbf{X}_x)}{\lambda_{\max}(\mathbf{X}^{\top}\mathbf{X})}\right)$, the rule of \eqref{eq:mp-rule} converges to the minimizer of \eqref{eq:mp-gvr-noc}.
\end{proposition}

The key point here is that albeit $\mathbf{X}$ is not symmetric, it is still positive-definite under the assumptions of Corollary~\ref{co:mp-X}. Hence, rule \eqref{eq:mp-rule} remains a valid descent direction for the unconstrained problem in \eqref{eq:mp-gvr-noc}.

\subsection{Constrained Reactive Support}\label{subsec:mp-con}
Reactive power injections are typically constrained in a feasible set $\mathcal{Q}$, and there may also be some reactive power compensation cost $c_2(\mathbf{q})$. In that case, the update in \eqref{eq:mp-rule} is not practical. Recall that for single-phase grids with constrained reactive resources, the original voltage magnitude deviation function $f_1(\mathbf{q})=\tfrac{1}{2}\|\mathbf{v}-v_0\mathbf{1}\|_2^2$ was replaced by a rotated Euclidean norm of the voltage deviations, namely $f_2(\mathbf{q})=\tfrac{1}{2}\|\mathbf{X}^{-1/2}(\mathbf{v}-v_0\mathbf{1})\|_2^2$, which resulted in localized updates. The choice of $f_1(\mathbf{q})$ would fail yielding localized solutions in multi-phase grids too. Although $\mathbf{X}$ in multi-phase grids is positive-definite, the lack of symmetry does not allow us to extend the approach with $f_2(\mathbf{q})$.

Nonetheless, let us study the behavior of the control rule of \eqref{eq:closed} under unbalanced conditions. Assume that the DG unit at each bus performs the control rule of \eqref{eq:sp-yn}--\eqref{eq:closed} that can be equivalently expressed as (cf.~\cite{SGC15})
\begin{subequations}\label{eq:mp-con}
\begin{align}
\mathbf{y}^t&=\mathbf{q}^t-\mu (\mathbf{v}^t-v_0\mathbf{1})\label{eq:mp-y}\\
\mathbf{q}^{t+1}&=\prox_{\mu c_2,\mathcal{Q}}[\mathbf{y}^t]\label{eq:mp-prox}
\end{align}
\end{subequations}
where the proximal operator is defined as
\begin{equation}\label{eq:proximal}
\prox_{\mu c_2,\mathcal{Q}}\left[\mathbf{y}\right]:=\arg\min_{\mathbf{w}\in \mathcal{Q}}~\mu c_2(\mathbf{w})+\tfrac{1}{2}\|\mathbf{w}-\mathbf{y}\|_2^2.
\end{equation}
Compared to the single-phase grid case, the major difference is that now $\mathbf{v}$ is related to $\mathbf{q}$ according to the model in \eqref{eq:mp-model}. The iterates produced by \eqref{eq:mp-con} satisfy:
\begin{align*}
\|\mathbf{q}^{t+1} - \mathbf{q}^{t}\|_2 &= \|\prox_{\mu c_2,\mathcal{Q}}[\mathbf{y}^{t}] - \prox_{\mu c_2,\mathcal{Q}}[\mathbf{y}^{t-1}]\|_2\\
&\leq \|\mathbf{y}^{t}-\mathbf{y}^{t-1}\|_2\\
&= \|\left(\mathbf{I} - \mu \mathbf{X}\right)\left(\mathbf{q}^{t}-\mathbf{q}^{t-1}\right)\|_2\\
&\leq \|\mathbf{I} - \mu \mathbf{X}\|_2 \|\mathbf{q}^{t}-\mathbf{q}^{t-1}\|_2.
\end{align*}
where the first inequality follows from the non-expansive property of the proximal operator (cf.~\cite[Prop.~5.1.8]{proximal}); the equality from \eqref{eq:mp-y}; and the last inequality from the definition of the maximum singular value. If $\mu$ is selected such that $\|\mathbf{I} - \mu \mathbf{X}\|_2<1$, then \eqref{eq:mp-con} constitutes a non-expansive mapping and it therefore converges to the equilibrium point defined by
\begin{equation}\label{eq:equi}
\mathbf{q}^*=\prox_{\mu c_2,\mathcal{Q}}[\mathbf{q}^*-\mu (\mathbf{v}(\mathbf{q}^*)-v_0\mathbf{1})]
\end{equation}
or, from the definition of the proximal operator in \eqref{eq:proximal}, by
\begin{equation*}
\mathbf{q}^*=\arg\min_{\mathbf{w}\in \mathcal{Q}}~\mu c_2(\mathbf{w})+\tfrac{1}{2}\|\mathbf{w}-[\mathbf{q}^*-\mu (\mathbf{v}(\mathbf{q}^*)-v_0\mathbf{1})]\|_2^2.
\end{equation*}
Vector $\mathbf{q}^*$ is thus defined as the minimizer of an optimization problem, and it cannot be expressed in closed form. Of course, the equilibrium point $\mathbf{q}^*$ does not necessarily coincide with the minimizer of any voltage regulation optimization problem. The next step size range guarantees $\|\mathbf{I} - \mu \mathbf{X}\|_2<1$, and therefore convergence of \eqref{eq:mp-con} (see the appendix for a proof):

\begin{proposition}\label{pro:mu-mp}
Let $\mathbf{U}\mathbf{\Lambda}\mathbf{U}^{\top}$ be the eigenvalue decomposition of $\mathbf{X}\mathbf{X}^{\top}$. If $\mu\in \left(0,\lambda_{\min}\left( \mathbf{\Lambda}^{-1/2}\mathbf{U}^{\top}(\mathbf{X}+\mathbf{X}^{\top})\mathbf{U}\mathbf{\Lambda}^{-1/2} \right)\right)$, then $\|\mathbf{I} - \mu \mathbf{X}\|_2<1$.
\end{proposition}

Practical distribution grids do not have all phases at all buses. The previous analysis carries over to such cases, if the related entries of $\mathbf{Z}_n$'s and the corresponding (re)active power injections are set to zero. For the eigendecompositions of $\mathbf{X}\mathbf{X}^{\top}$ and $\mathbf{X}+\mathbf{X}^{\top}$, their rows and columns related to non-existing bus-phase pairs are simply removed.

\section{Numerical Tests}\label{sec:tests}
The voltage regulation schemes presented earlier are evaluated using the IEEE 13-bus feeder, the IEEE 123-bus feeder, and a Southern California Edison (SCE) 47-bus feeder \cite{GLTL12}, \cite{PSTCA}. Solar generation data from the Smart* project and from August 24, 2011 are used~\cite{SmartStar}. Unless otherwise stated, active power injections are kept fixed over the reactive control period, and PV reactive injections are initialized to zero.

\begin{figure}[t]
\centering
\includegraphics[scale=0.6]{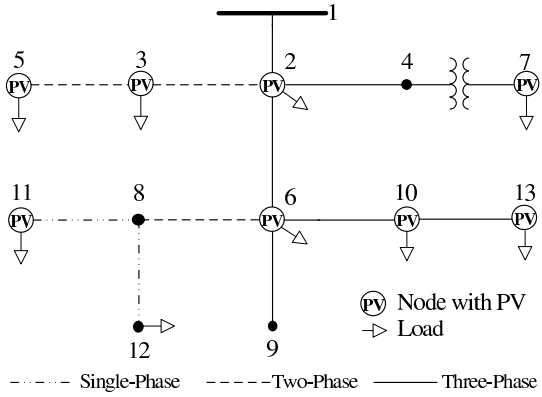}
\caption{IEEE 13-bus feeder~\protect{\cite{PSTCA}}.}
\label{fig:13bus}
\end{figure}


\subsection{Single-Phase Grids}\label{subsec:sp-sims}
Single-phase grids are obtained upon modifying multi-phase grids as in \cite{SGC15}. The global minimizer of \eqref{eq:gvr} is obtained using MATLAB and it serves as a benchmark. The first experiment simulates an over-voltage scenario on the IEEE 13-bus grid depicted in Fig. \ref{fig:13bus}. The IEEE 13-bus grid exhibits $\kappa(\mathbf{X})=716$. Simulating a severe over-voltage violation in the transmission network, the feeder voltage magnitude is set to 1.07 p.u., while the voltage regulator is removed from the system. A 52\% PV penetration level is assumed for all buses apart from buses 4, 8, 9, and 12. Loads are fixed to 80\% of their peak value, and reactive power marginal costs are set to $c_n=0.0125\cent$/kVar\& h for all $n$.

\begin{figure}[t]
\centering
\includegraphics[scale=0.35]{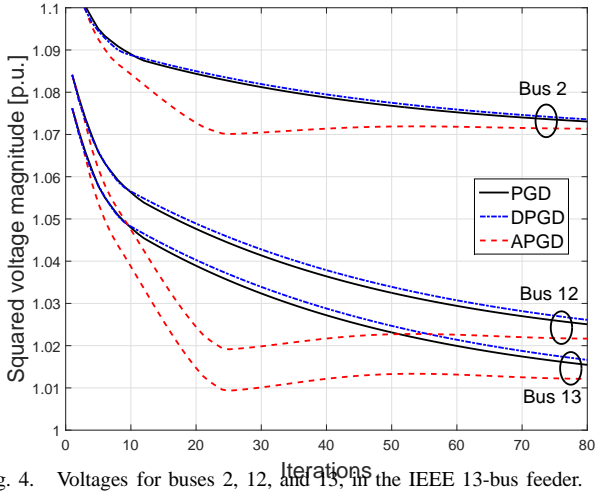}
\vspace*{-3em}
\caption{Voltages for buses 2, 12, and 13, in the IEEE 13-bus feeder.}
\label{fig:13v}
\end{figure}

Three reactive power control rules are tested: (i) the proximal gradient descent (PGD) of \eqref{eq:closed}; (ii) the proximal gradient descent with diagonal scaling (DPGD) in \eqref{eq:closed2}; and the accelerated proximal gradient descent (APGD) of \eqref{eq:closed3}. The squared voltage magnitudes obtained at three buses are illustrated in Fig.~\ref{fig:13v}. For all control rules, the step size is conservatively set to $\mu=0.1/\lambda_{\max}(\mathbf{X})$. Figure~\ref{fig:13v} demonstrates that APGD has an obvious four-fold speedup advantage over PGD and DPGD. Diagonal scaling does not exhibit any convergence rate advantage over PGD. The latter could be explained by the fact that the diagonal entries of $\mathbf{X}$ have similar values; hence, matrices $\mathbf{X}$ and $\mathbf{D}^{1/2}\mathbf{X}\mathbf{D}^{1/2}$ with $\mathbf{D}=\diag(\diag(\mathbf{X}))$ have similar condition numbers. On a different note and as expected, the minimizer of \eqref{eq:gvr-f2} does not guarantee small voltage magnitude deviations across all buses: Buses 12-13 exhibit small deviations, yet bus 2 converges to deviating from nominal by 1.03.

\begin{figure}[t]
\centering
\includegraphics[scale=0.35]{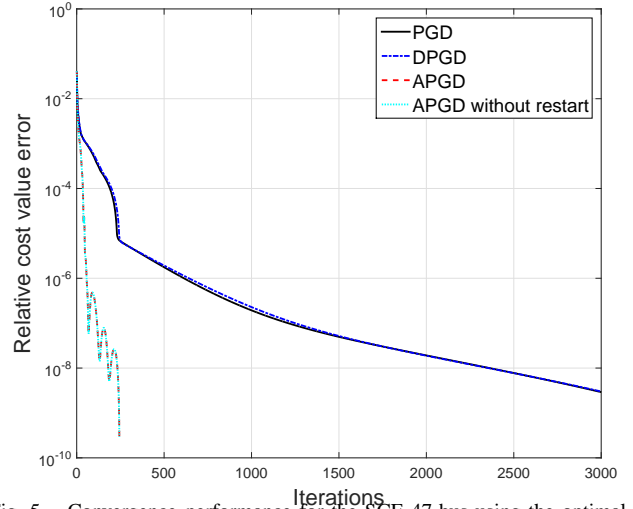}
\vspace*{-3em}
\caption{Convergence performance for the SCE 47-bus using the optimal $\mu$.}
\label{fig:47cost}
\end{figure}

The second experiment involves the SCE 47-bus grid with $\kappa (\mathbf{X})=16,470$~\cite{GLTL12}. The capacitor located on bus 32 is ignored. Loads are set to 80\% of their peak value with a power factor of 0.8. Five PV generators generating 60\% of their capacity are located on buses 13, 17, 19, 23, 24. Distributed PVs with 50\% penetration level are further installed on buses 11, 12, 14, 22, 25, 33, 38, 39, 41. The relative cost value error attained by \eqref{eq:gvr-f2} using the optimal $\mu=1/\lambda_{\max}(\mathbf{X})$ is depicted in Fig.~\ref{fig:47cost}. Apparently, the novel scheme converges at least six to ten times faster than its alternatives. 

\begin{figure}[t]
\centering
\includegraphics[scale=0.35]{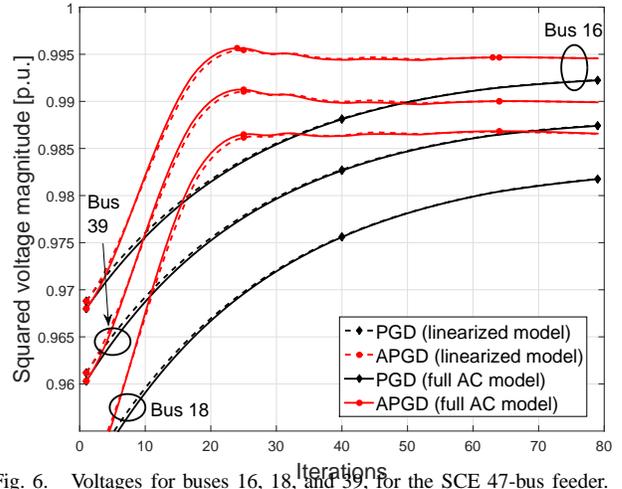}
\vspace*{-3em}
\caption{Voltages for buses 16, 18, and 39, for the SCE 47-bus feeder.}
\label{fig:47v}
\end{figure}

We further tested the accuracy of the linearized model over the full AC model calculated using the forward-backward algorithm~\cite{Kersting}. Voltage magnitudes obtained from PGD and APGD  with $\mu=0.1/\lambda_{\max}(\mathbf{X})$ are presented in Fig.~\ref{fig:47v}. The curves suggest that the linearized model of \eqref{eq:sp-model} offers a good approximation.

\begin{figure}[t]
\centering
\includegraphics[scale=0.35]{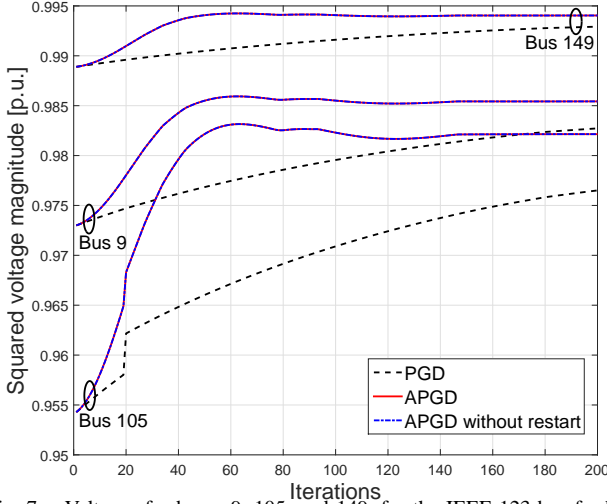}
\vspace*{-3em}
\caption{Voltages for buses 9, 105, and 149, for the IEEE 123-bus feeder.}
\label{fig:123v}
\end{figure}

To evaluate the control schemes on more realistic conditions, the third experiment   uses the IEEE 123-bus grid and it also entails a topology reconfiguration: After twenty control periods (algorithm iterations), the switch between buses 97 and 197 opens, while the switch between buses 151 and 300 closes. Renewable (PV) generation units are located on buses 32, 51, 64, 76, 80, 93, and 114, with capacities 60, 60, 120, 80, 30, 100, and 80 kVA, respectively. The condition number for this feeder is $\kappa(\mathbf{X})=20,677$. Figure~\ref{fig:123v} presents the squared voltage magnitudes over three representative buses and for $\mu=0.1/\lambda_{\max}(\mathbf{X})$. Three observations are in order. First, note that the convergence guarantees for all three control schemes hold for any feasible initialization point. Theoretically, a topology change could potentially delay APGD since the parameters $\beta_t$ in \eqref{eq:closed3} are time-increasing. According to the curves, the topology change does not affect significantly the convergence rate of any of the algorithms. Second, the APGD scheme exhibits superior convergence properties over the PGD rule. Third, compared to the experiments on the IEEE 13-bus and the SCE 47-bus feeders, the longer convergence period can be attributed to the larger size of the feeder.

\subsection{Multi-Phase Grids}\label{subsec:mp-sims}
\begin{figure}[t]
\centering
\includegraphics[scale=0.35]{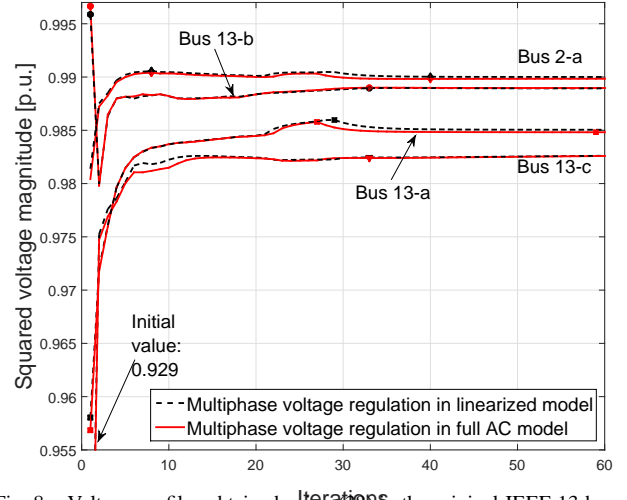}
\vspace*{-3em}
\caption{Voltage profiles obtained using \eqref{eq:mp-con} in the original IEEE 13-bus grid.}
\label{fig:mp13v}
\end{figure}

To evaluate the findings of Section~\ref{sec:mp-alg}, the control rule described by the iterations in \eqref{eq:mp-con} is tested on the multi-phase IEEE 13-bus system. Loads and PV penetration are selected as in the single-phase experiment presented earlier. The feeder voltage magnitude is fixed to 1, while the step size is set to $\mu = \lambda_{\min}\left( \mathbf{\Lambda}^{-1/2}\mathbf{U}^{\top}(\mathbf{X}+\mathbf{X}^{\top})\mathbf{U}\mathbf{\Lambda}^{-1/2} \right)$. Squared voltage magnitude profiles obtained from \eqref{eq:mp-con} are plotted in Fig.~\ref{fig:mp13v}. Voltages are calculated using both the linearized model and the full AC model at every iteration. Verifying the findings of \cite{PSCC14}, the curves indicate that the approximation is quite precise. The control rule converges within 40 iterations. Tests conducted with smaller step sizes exhibit slower convergence, whereas the scheme diverges for a step size $\mu \geq 3.1\lambda_{\min}\left( \mathbf{\Lambda}^{-1/2}\mathbf{U}^{\top}(\mathbf{X}+\mathbf{X}^{\top})\mathbf{U}\mathbf{\Lambda}^{-1/2} \right)$.

\section{Conclusions}\label{sec:conclusions}
To derive communication-free solutions, voltage regulation goals were lowered to penalizing large voltage magnitude deviations. For balanced distribution grids, localized (un)constrained schemes were analyzed, while a control scheme based on proximal gradient descent was developed. Its computational complexity is comparable to that of the control rule suggested by the IEEE 1547.8 standard~\cite{IEEE1547}, yet it enjoys precise convergence guarantees. Adding memory to the control rule further yielded a significantly faster voltage regulation scheme. For unbalanced distribution grids, a linear approximation model revealed a counter-intuitive inter-phase coupling. Based on the properties of the involved matrices, the developed reactive power control rule converges to a well-defined equilibrium point. Determining whether the latter point corresponds to the minimizer of a particular optimization problem is a challenging open question. Numerical tests on benchmark feeders indicated the superiority of the accelerated scheme over diagonal scaling, and the resiliency of all novel schemes against topology reconfigurations. Incorporating active power curtailment and performing local per-bus processing across phases constitute interesting future research directions.

\appendix\label{sec:appendix}
\begin{IEEEproof}[Proof of Proposition~\ref{pro:B}]
Collecting \eqref{eq:m-amodelv} for all $n$ yields
$\mathbf{v}_{\pi} - \mathbf{v} = 2\real\left[ \bdiag(\{\tilde{\mathbf{Z}}_n\}) \check{\mathbf{S}} \right]$,
or via the permutations of \eqref{eq:permute}, as
\begin{equation}\label{eq:sproof1b}
\mathbf{T}\left(\check{\mathbf{v}}_{\pi} - \check{\mathbf{v}} \right)= 2\real\left[ \bdiag(\{\tilde{\mathbf{Z}}_n\}) \mathbf{T}\mathbf{S} \right].
\end{equation}

Focus first on the LHS of \eqref{eq:sproof1b}. Observe that voltage drops between adjacent buses can be alternatively expressed as $\check{\mathbf{v}}_{\pi,\phi}- \check{\mathbf{v}}_{\phi}=\mathbf{A}\check{\mathbf{v}}_{\phi} + v_0\mathbf{a}_0$ for $\phi\in\{a,b,c\}$. Stacking the latter equations across all phases yields
\begin{equation}\label{eq:sproof2}
\check{\mathbf{v}}_{\pi} - \check{\mathbf{v}} =  (\mathbf{I}_3\otimes \mathbf{A}) \check{\mathbf{v}} + v_0\mathbf{1}_3\otimes \mathbf{a}_0. 
\end{equation}

Regarding the RHS of \eqref{eq:sproof1b}, recall that flows are decoupled across phases. The grid topology and \eqref{eq:m-amodels} dictate that $\check{\mathbf{s}}_{\phi}=\mathbf{A}^{\top} \check{\mathbf{S}}_{\phi}$, or equivalently, $\check{\mathbf{S}}_{\phi}=-\mathbf{F}^{\top} \check{\mathbf{s}}_{\phi}$ for all $\phi$. Stacking flows across all phases yields
\begin{equation}\label{eq:sproof3}
\check{\mathbf{S}} = -(\mathbf{I}_3\otimes \mathbf{F}^{\top})\check{\mathbf{s}}.
\end{equation}

Plugging \eqref{eq:sproof2}--\eqref{eq:sproof3} into \eqref{eq:sproof1b}, and solving for $\check{\mathbf{v}}$ results in
\begin{align}\label{eq:sproof4}
\check{\mathbf{v}} =& -2\real\left[ (\mathbf{I}_3\otimes \mathbf{A})^{-1} \mathbf{T}^{-1} \bdiag (\{\mathbf{Z}_n\}) \mathbf{T} (\mathbf{I}_3 \otimes \mathbf{F}^{\top}) \check{\mathbf{s}} \right]\nonumber\\
& -  v_0 (\mathbf{I}_3\otimes \mathbf{A})^{-1} (\mathbf{1}_3\otimes \mathbf{a}_0).
\end{align}
Using the facts $\mathbf{A}^{-1}=-\mathbf{F}$, $\mathbf{T}^{-1}=\mathbf{T}^{\top}$, $\mathbf{F}\mathbf{a}_0=\mathbf{1}_N$, and properties of the Kronecker product, \eqref{eq:sproof4} becomes
\begin{align}\label{eq:sproof5}
\check{\mathbf{v}} =& 2\real\left[ (\mathbf{I}_3\otimes \mathbf{F}) \mathbf{T}^{\top} \bdiag(\{\tilde{\mathbf{Z}}_n\}) \mathbf{T} (\mathbf{I}_3 \otimes \mathbf{F}^{\top}) \check{\mathbf{s}} \right]\nonumber +  v_0\mathbf{1}_{3N}.
\end{align}
Substituting $\check{\mathbf{s}}=\mathbf{T}^{\top}\mathbf{s}$ and $\check{\mathbf{v}}=\mathbf{T}^{-1}\mathbf{v}$ proves the claim.
\end{IEEEproof}

\begin{IEEEproof}[Proof of Proposition~\ref{pro:sp-descent}]
The claim is an application of [7, Prop.~2.1]. The Lipschitz constant of $\nabla f_1(\mathbf{q})$ is $\lambda_{\max}^2(\mathbf{X})$. Observe also that $\|\mathbf{g}^t\|_2^2=(\mathbf{v}^t-v_0\mathbf{1})^{\top}\mathbf{X}\mathbf{X}^{\top}(\mathbf{v}^t-v_0\mathbf{1}) \leq \lambda_{\max}^2(\mathbf{X})\|\mathbf{v}^t-v_0\mathbf{1}\|_2^2$, or
\begin{align*}
\|\mathbf{v}^t-v_0\mathbf{1}\|_2 \geq \frac{1}{\lambda_{\max}(\mathbf{X})} \|\mathbf{g}^t\|_2.
\end{align*}
Note further that
$(\mathbf{v}^t-v_0\mathbf{1})^{\top}\mathbf{g}^t=(\mathbf{v}^t-v_0\mathbf{1})^{\top}\mathbf{X}(\mathbf{v}^t-v_0\mathbf{1})\geq \lambda_{\min}(\mathbf{X})\|\mathbf{v}^t-v_0\mathbf{1}\|_2^2$,
implying that $(\mathbf{v}^t-v_0\mathbf{1})$ is a descent direction for $f_1(\mathbf{q}^t)$ -- although not the steepest one.
\end{IEEEproof}

\begin{IEEEproof}[Proof for Proposition~\ref{pro:mp-descent}]
As in Prop.~\ref{pro:sp-descent}, showing Prop.~\ref{pro:mp-descent} relies on an application of~[7, Prop.~2.1]. In this case, the Lipschitz constant for $\nabla f_1(\mathbf{q})$ is $\lambda_{\max}(\mathbf{X}^{\top}\mathbf{X})$, and it also holds that $\|\mathbf{v}_t-v_0\mathbf{1}\|_2\geq \lambda_{\min}^{-1/2}(\mathbf{X}^{\top}\mathbf{X})\|\nabla f_1(\mathbf{q}^t)\|_2$. The critical point in the multi-phase grid case though is that the positive-definiteness of $\mathbf{X}_x$ guarantees that $(\mathbf{v}^t-v_0\mathbf{1})$ is a descent direction, since
$(\mathbf{v}^t-v_0\mathbf{1})^{\top}\nabla f_1(\mathbf{q}^t)= (\mathbf{v}^t-v_0\mathbf{1})^{\top}\mathbf{X}^{\top}(\mathbf{v}^t-v_0\mathbf{1})
= (\mathbf{v}^t-v_0\mathbf{1})^{\top}\mathbf{X}_x(\mathbf{v}^t-v_0\mathbf{1})\leq \lambda_{\min}(\mathbf{X}_x)\|\mathbf{v}^t-v_0\mathbf{1}\|_2^2.$
\end{IEEEproof}

\begin{IEEEproof}[Proof of Proposition~\ref{pro:mu-mp}]
By definition, it holds that
\begin{equation*}
\|\mathbf{I} - \mu \mathbf{X}\|_2^2=\lambda_{\max}\left((\mathbf{I}-\mu \mathbf{X})^{\top}(\mathbf{I}-\mu \mathbf{X})\right) = \lambda_{\max}\left(\mathbf{I}-\mu \mathbf{X}_{\mu}\right)
\end{equation*}
where $\mathbf{X}_{\mu}:=\mathbf{X}+\mathbf{X}^{\top}-\mu\mathbf{X}\mathbf{X}^{\top}$. Guaranteeing $\|\mathbf{I} - \mu \mathbf{X}\|_2<1$ is equivalent to satisfying $ \lambda_{\max}\left(\mathbf{I}-\mu \mathbf{X}_{\mu}\right)<1$, or, simply $1-\mu\lambda_{\min}\left(\mathbf{X}_{\mu}\right)<1$. Granted that $\mu>0$, the latter is equivalent to ensuring $\mathbf{X}_{\mu}$ to be a positive-definite matrix, i.e., $\mathbf{X}+\mathbf{X}^{\top}\succ \mu\mathbf{X}\mathbf{X}^{\top}$. It can be easily verified that pre/post-multiplying the aforementioned linear matrix inequality by $\mathbf{\Lambda}^{-1/2}\mathbf{U}^{\top}$/$\mathbf{U}\mathbf{\Lambda}^{-1/2}$ yields the condition imposed on $\mu$ by Proposition~\ref{pro:mu-mp}.
\end{IEEEproof}

\bibliographystyle{IEEEtranS}
\bibliography{myabrv,power}
\end{document}